\documentclass[11pt]{article}
\usepackage{moriond,epsfig}
\usepackage{amsmath}
\usepackage{subfig}

\def\be{\begin{equation}}
\def\ee{\end{equation}}
\def\bea{\begin{eqnarray}}
\def\eea{\end{eqnarray}}

\input{ogbabarsym}

\begin{document}
\vspace*{4cm}
\title{Baryonic \B decays at \babar}

\author{ O. GRUENBERG (for the \babar\ collaboration) }

\address{University of Rostock, Institute for Physics,\\
Uniplatz 3, 18055 Rostock, Germany}

\maketitle\abstracts
{We report on the analyses of the baryonic $\B$ decays $\mydecII$ and $\mydec$. 
The underlying data sample consists of $470 \times 10^6$ \BB pairs generated in 
the process \epem\to\upsbb and collected with the \babar\ detector at the \pep2 \ 
storage ring at SLAC. 
We find $\BR(\mydecII)\cdot\BR(\Lcp\to\proton\Km\pip)/5\%<6.2\cdot10^{-6}\ @\ 90\,\%\ CL$
and $\BR(\mydec)=(2.98\pm0.16_{\stat}\pm0.15_{\syst}\pm0.77_{(\rm\Lambda_c)})\times10^{-4}$,
where the last error is due to the uncertainty in $\BR(\Lcp\to\proton\Km\pip)$.
The data suggest the existence of resonant subchannels $\Bub\to\LcpsI\antiproton\pim$
and, possibly, $\mydecresA$. We see unexplained structures in $m(\Scpp\pim\pim)$ at $3.25\gevcc$, 
$3.8\gevcc$, and $4.2\gevcc$.}

\section{Introduction}

Approximately $7\,\%$ \cite{ref:PDG} of all \B mesons have baryons among their
decay products. This is a substantial fraction that justifies further investigations
which may allow better understanding of baryon production in \B decays and, more 
generally, hadron fragmentation into baryons. The measurement and comparison of exclusive 
branching fractions of baryonic \B decays as well as systematic studies on the dynamic of the decay, 
i.e. the fraction of resonant subchannels, is a direct way to study the mechanisms of baryonization.
In the following, we present the results of two recently completed \babar\ analyses of the decays 
\mydec and \mydecII \cite{footnote}.

\section{$\boldsymbol{\mydec}$}

The decay \mydec is a resonant subchannel of the five body final state \mydecF, which has, 
until now, the largest known branching fraction among all baryonic \B decays and hence is 
a good starting point for further investigations.\\

\subsection{Reconstruction}

\begin{figure}[ht]
\begin{center}
	\includegraphics[width=0.5\textwidth]{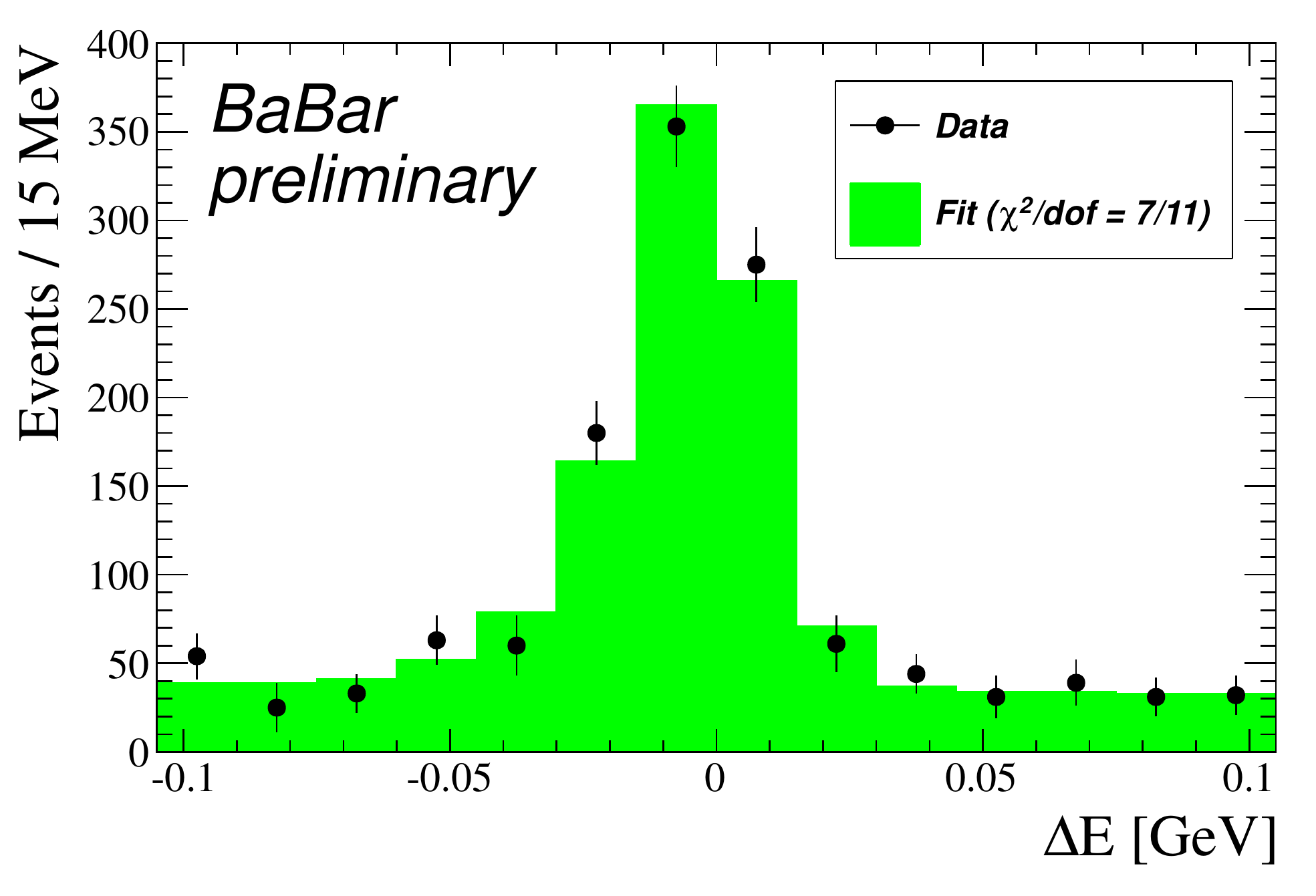}
	\caption{The distribution of \DeltaE from the \babar\ data.}
	\label{fig:1}
\end{center}
\end{figure}

\noindent
We reconstruct the decay in the subchannel \Scpp\to\Lcp\pip, and \Lcp\to\proton\Km\pip.
For the signal selection we use the missing energy of the \B candidate in the \epem rest 
frame: $\DeltaE=\sqrt{{\rm E_{B}^{2*}}-\sqrt{s}/2}$.
Figure \ref{fig:1} shows the distribution of \DeltaE from the sample of 
reconstructed \B events in data after selections for background suppression.
From a fit we find $787\pm43$ signal events. The reconstruction efficiency is 
$(11.3\pm0.2_{\stat})\%$. The branching fraction is
$\BR(\mydec)=(2.98\pm0.16_{\stat}\pm0.15_{\syst}\pm0.77_{(\rm \Lambda_c)})\times10^{-4}$.

\subsection{Resonant subchannels}

\begin{figure}[ht]
\begin{center}
\subfloat[]
{\includegraphics[width=0.48\textwidth]{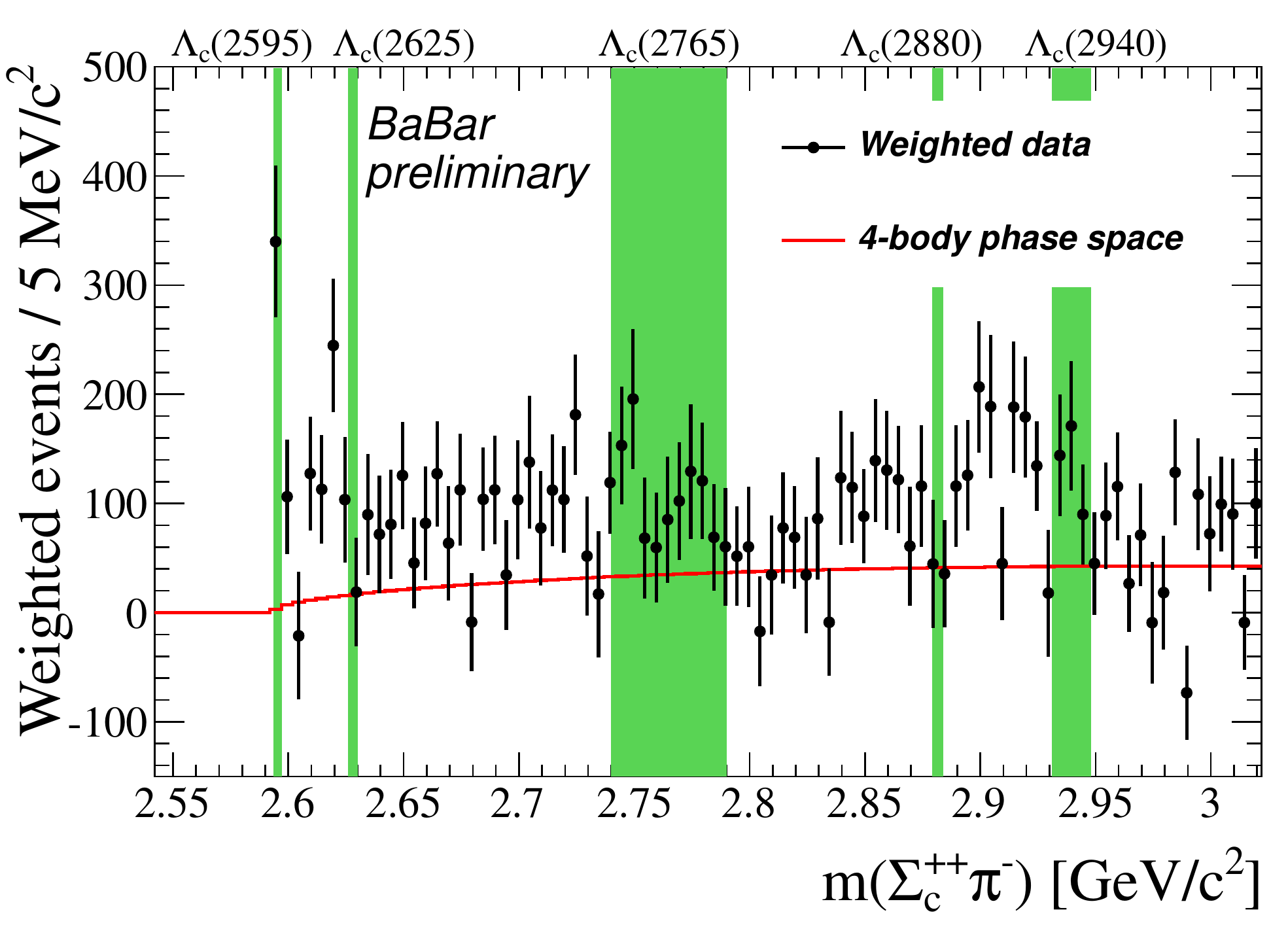}}
\rule{0.5em}{0em}
\subfloat[]
{\includegraphics[width=0.50\textwidth]{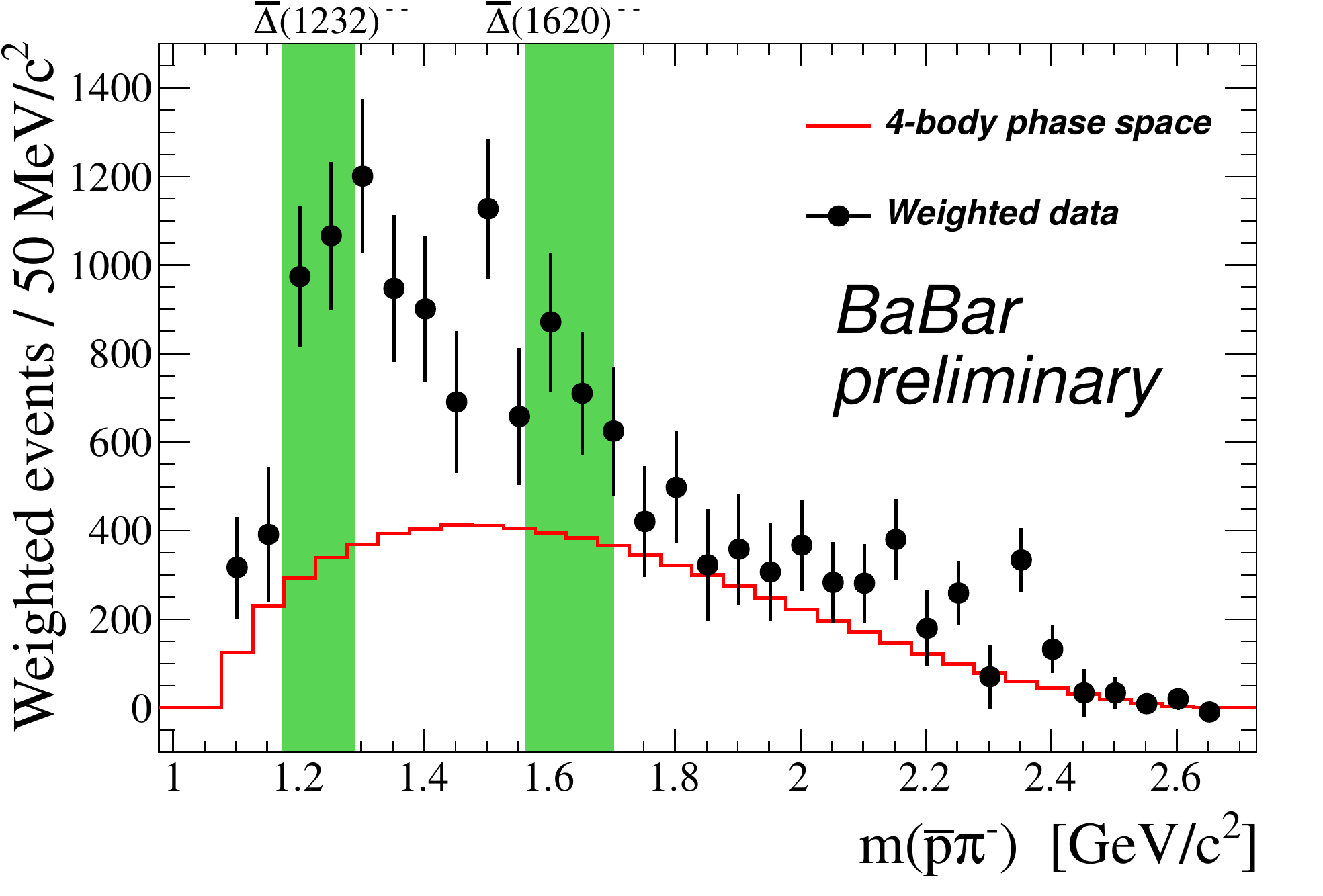}}
\caption{The distribution of m(\Scpp\pip) and m(\proton\pim) from \babar\ data and four-body PS.}
\label{fig:2}
\end{center}
\end{figure}

We see large deviations between data and the prediction of four-body phase space (PS)
in the two-body and three-body masses of the \B daughters. These deviations may
be attributed to the resonant intermediate states $\Lcps\to\Scpp\pim$ and 
$\Deltabar^{--}\to\antiproton\pim$.\\ 
Figure \ref{fig:2}(a) shows the invariant mass distribution of \Scpp\pim after a 
sideband subtraction in \DeltaE and efficiency correction. The large number of events at 
the threshold is compatible with the existance of the resonance $\Lcp(2595)^{+}$. 
There are no significant signals for other \Lcps resonances.\\
Figure \ref{fig:2}(b) shows the invariant mass distribution of $\proton\pim$ after a sideband
subtraction in \DeltaE and efficiency correction. The differences between data and PS in the 
range of $m(\antiproton\pim)\in(1.2,1.7)\gevcc$ could be due to the existance of the 
resonances $\Deltabar^{--}(1232,1600,1620)$.\\

\begin{figure}[ht]
\begin{center}
\subfloat[]{\includegraphics[width=0.5\textwidth]{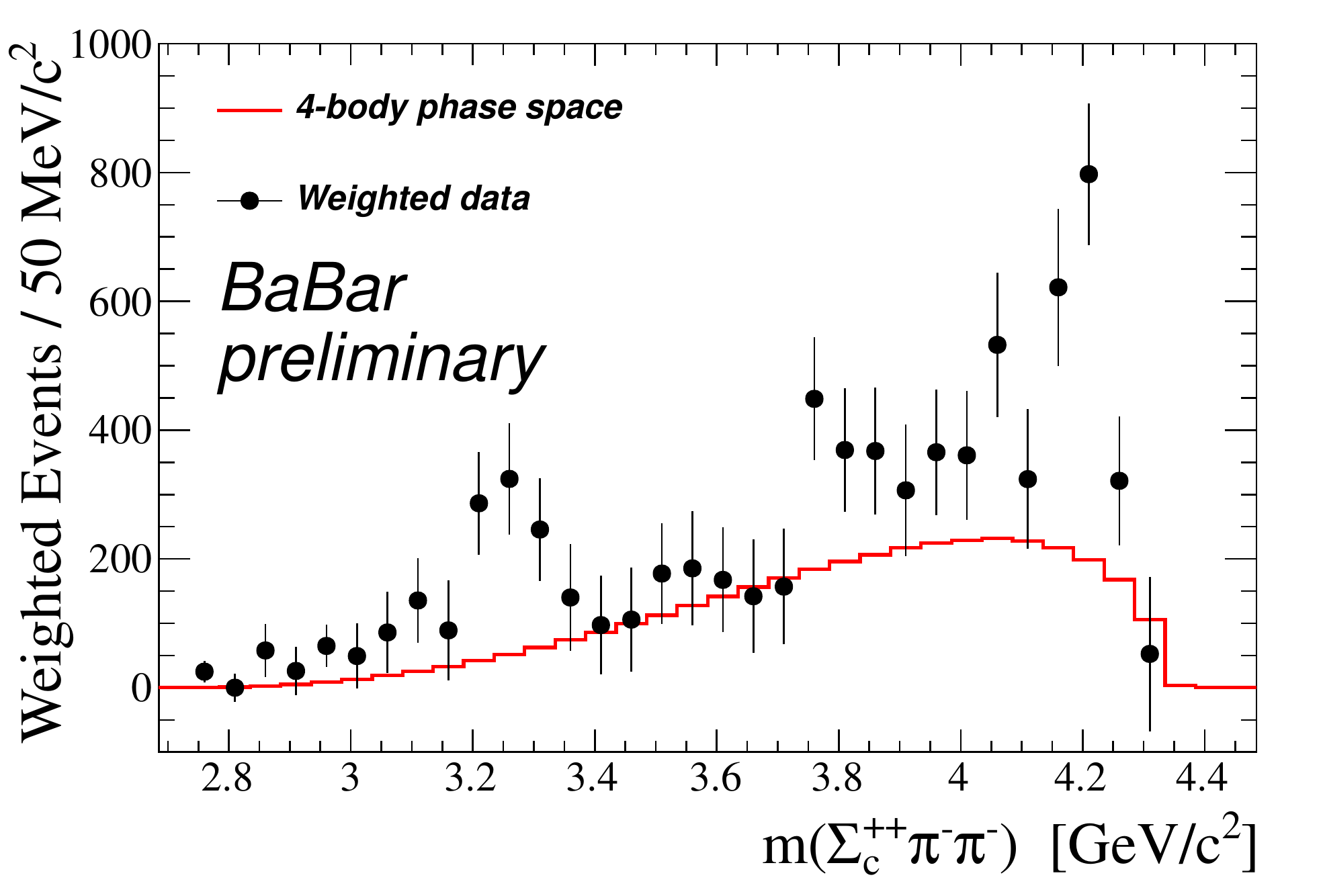}}
\subfloat[]{\includegraphics[width=0.5\textwidth]{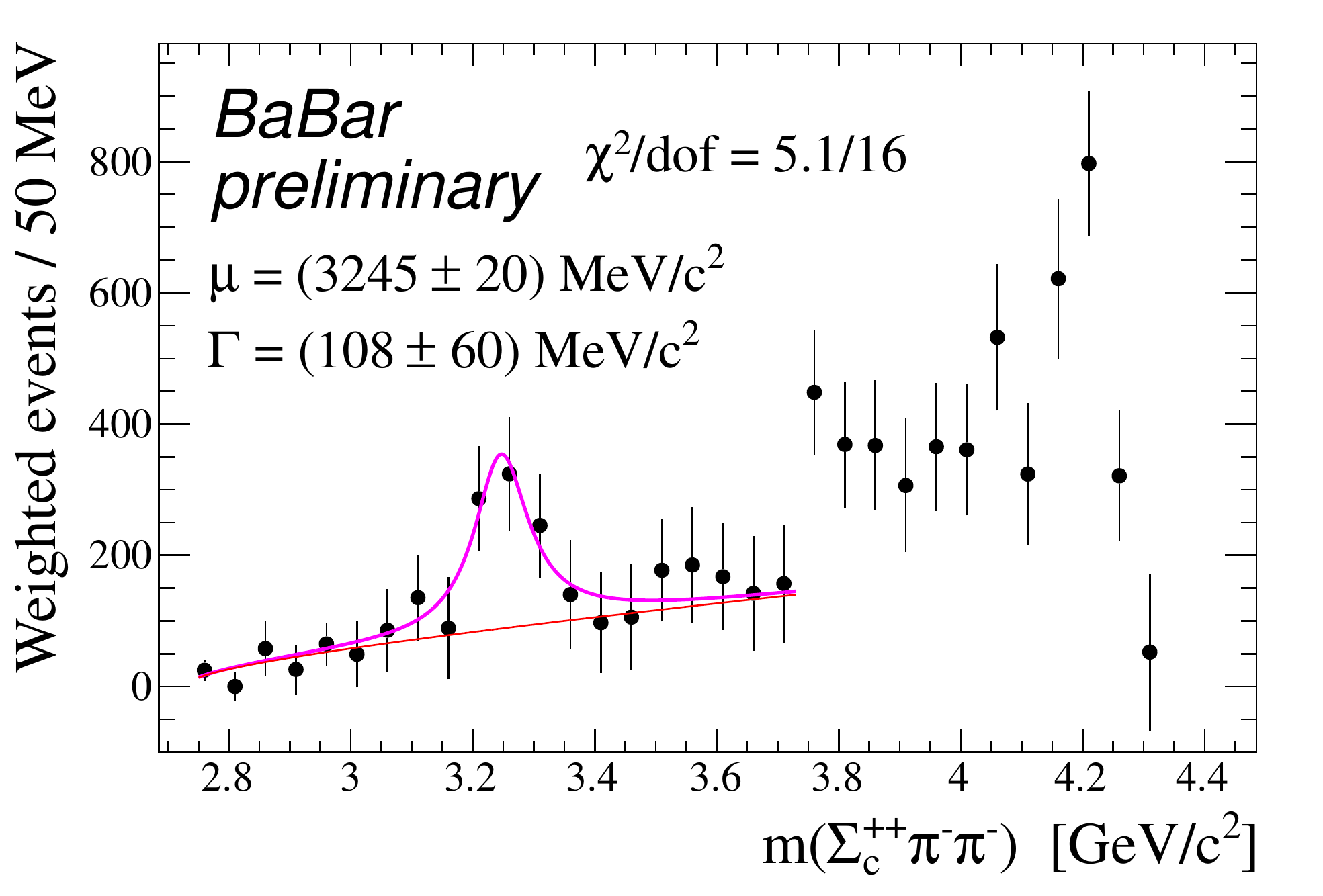}}
\caption{The distribution of m(\Scpp\pim\pim) from \babar\ data and four-body PS.}
\label{fig:3}
\end{center}
\end{figure}

\noindent
Figure \ref{fig:3}(a) shows the invariant mass distribution of $\Scpp\pim\pim$ after 
a sideband subtraction in \DeltaE and efficiency correction. We see unexplained structures 
at $3.25\gevcc$, $3.8\gevcc$, and $4.2\gevcc$.
In figure \ref{fig:3}(b) we present the result of a fit in the range 
$m(\Scpp\pim\pim)=2.750\ldots3.725\gevcc$. We choose an ad-hoc parametrization that consists of a 
Breit-Wigner function with two parameters (width: $\Gamma$, mean: $\mu$) for the signal and a
two-body phase space distribution with the parameters $m_1=m(\Scpp)$ and $m_2=2\cdot m(\pim)$ 
for the background. The fitted parameters are $\mu=(3245\pm20_{\stat})\mevcc$ and 
$\Gamma=(108\pm60_{\stat})\mevcc$.

\subsection{Conclusion}

\noindent
Comparing the branching fractions $\BR(\mydec)=(2.98\pm0.8)\times10^{-4}$
and $\BR(\mydecS)=(4.4 \pm 1.7)\times10^{-4}$ \cite{ref:CLEO} one finds that 
the decay \mydecS is $50\%$ more frequent. This could be due to a number of additional 
resonant subchannels that contribute to \mydecS, i.e. $\Bub\to\Scz\Nbar\pim$ and 
$\Bub\to\Scz\antiproton\rho^{0}$, and would indicate the importance of resonant subchannels 
in baryonic \B decays. Furthermore, the combined branching fraction of \mydec and \mydecS 
makes about $30\%$ of the branching fraction of the five body decay 
$\BR(\mydecF)=(22.5\pm 6.8)\times10^{-4}$ \cite{ref:CLEO}, which also stresses the 
large impact of intermediate states.

\section{\mydecII}

The decay \mydecII is one of a few allowed \B decays with a \b\to\c transition and four 
baryons in the final state. It is closely connected to \thdec 
($\BR=(1.12 \pm 0.32)\times10^{-3}$ \cite{ref:Belle2}) and \mydecF, which have similar 
quark contents and the (so far) largest measured branching fractions among the baryonic 
\B decays with a \Lcp in the final state. The main differences between the sought decay 
and the other two decay channels are the absence of possible resonant subchannels and 
the much smaller phase space $(Q(\mydecII)=176\mevcc, Q(\thdec)=1776\mevcc$ 
with $Q=m({\rm mother})-\sum m({\rm daughter})$.
The latter may favour the decay \mydecII, in that baryons are more likely to form when quarks are 
close to each other in momentum space \cite{Rosner:2003bm}, \cite{Suzuki:2006nn}. An example of 
this behavior is the ratio of $\BR(\Bub\to\Lcp\aLcp\Km)/\BR(\mjdecb)\approx3$ \cite{ref:PDG}, 
preferring the more massive final state that mainly differs by the size of phasespace since 
$|V_{\c\s}|\approx|V_{\u\d}|$. On the other hand the decay \mydecII may be suppressed by the 
fact that it does not have resonant subchannels which could play an important role for baryonic 
\B decays, i.e. $\BR(\thdec)_{\rm resonant}/\BR(\thdec)\approx40\%$ \cite{ref:PDG}.
The size of the branching fraction may allow to balance the relevance of resonant subchannels 
against momentum space in baryonic \B decays.\\

\subsection{Reconstruction}

\begin{figure}[ht]
\begin{center}
	\includegraphics[width=0.6\textwidth]{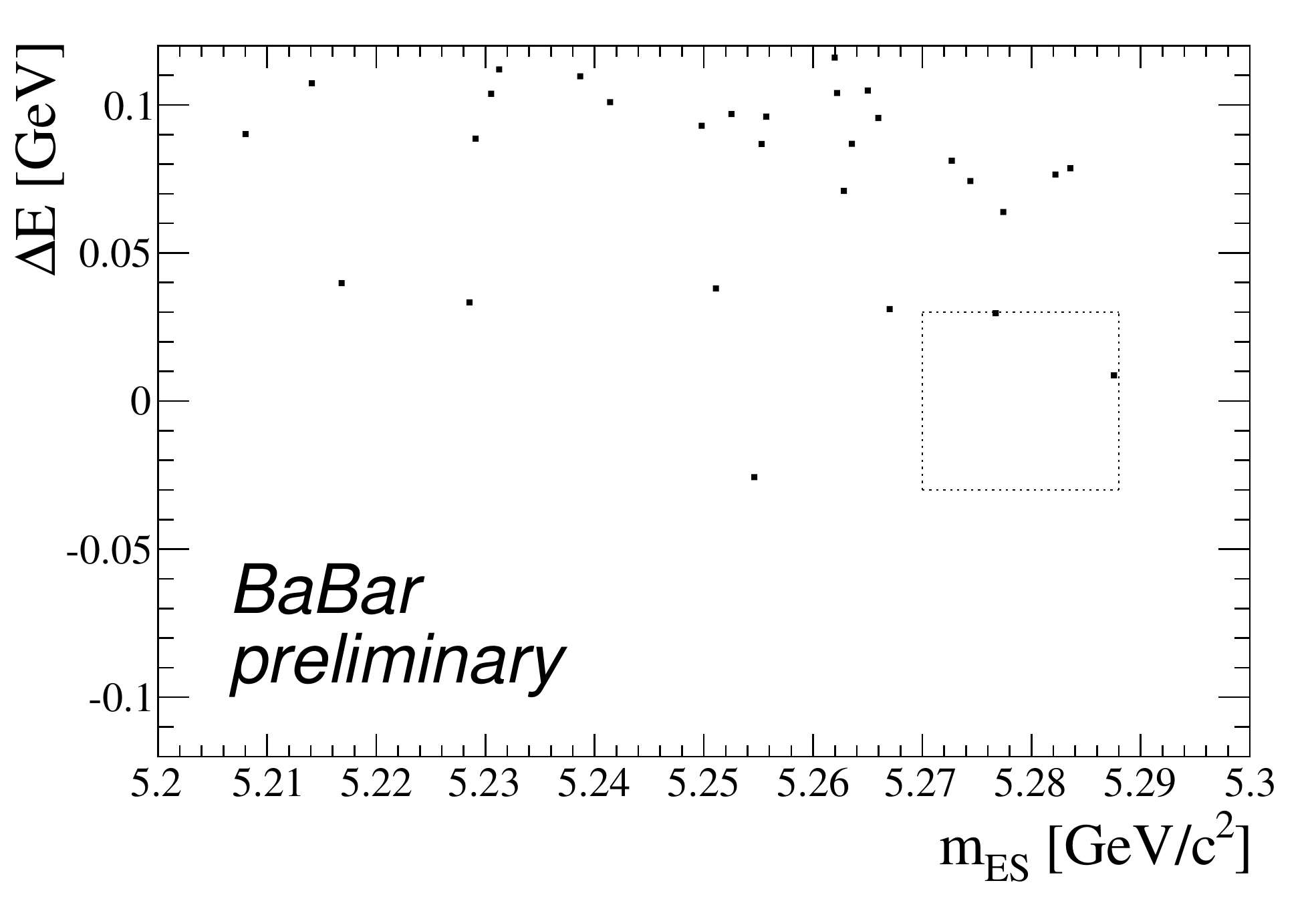}
	\caption{The distribution of \DeltaE vs. \mes from the \babar\ data.}
	\label{fig:4}
\end{center}
\end{figure}

\noindent
We reconstruct the decay \mydecII in the subchannel \Lcp\to\proton\Km\pip. Besides \DeltaE,
we use the energy substituted mass \mes of the \B candidate for the signal selection.
In a simplified form, it can be written as 
$\mes\prime=\sqrt{\left(\sqrt{s}/2\right)^{2}-|\vec{p}_{\rm B}^{\,*}|^{2}}$,
where $\vec{p}_{\rm B}^{\,*}$ is the momentum of the \B candidate in the \epem rest frame.
The complete formular of \mes also takes into account the asymmetric energies of \ep and \en.
\mes is centered at the true \B mass for correctly reconstructed \B decays.\\
Figure \ref{fig:4} shows the distribution of \DeltaE vs. \mes with a selection in $m_{\proton\Km\pip}$ 
for background suppression. There are two \B candidates within a signal window that is chosen on the 
basis of an analysis of simulated signal events. The efficiency in this range is 
$\eps = (3.66\pm0.03_{\stat})\,\%$. For background estimation we analyze sidebands in 
$m_{\proton\Km\pip}$ and \mes from the data sample as well as a set of simulated \babar\ 
events and find no reliable prediction due to large systematic uncertainties.
Therefore we calculate a conservative upper limit by taking the two \B candidates as signal.
In addition, we exclude the large uncertainty of $\BR(\Lcp\to\proton\Km\pip)=(5.0\pm1.3)\,\%$ \cite{ref:PDG}.
Consequently, we determine:

\vspace{-0.25em}

\begin{equation}
\mybrII \cdot \dfrac{\BR(\Lcp\to\proton\Km\pip)}{5\,\%}
< \dfrac{N_{up}}{\varepsilon\cdot N_{\B}\cdot 5\,\%} = 6.2\cdot10^{-6} \quad @\ CL=90\%
\end{equation}

\vspace{-0.25em}

\noindent
As a result we find that \mybrII is at least two orders of a magnitude smaller than \BR(\thdec) and
\BR(\mydecF) @ $CL=90\%$. This could indicate, that the phase space of \mydecII is too small to favor 
baryonisation of the quarks and thus increase the decay rate.
Furthermore, the nonappearance of resonant subchannels may additionally affect the branching fraction.




\begin{thebibliography}{99}
\bibitem{footnote}
Throughout this paper, all decay modes represent that mode and its charge conjugate.
\end{thebibliography}
\end{document}